# AN IMPROVED WATCHDOG TECHNIQUE BASED ON POWER-AWARE HIERARCHICAL DESIGN FOR IDS IN WIRELESS SENSOR NETWORKS

A. Forootaninia[1] and M. B. Ghaznavi-Ghoushchi [2*]

[1] Department of IT, Tehran University, Kish International Campus, Iran
`forootaninia@ut.ac.ir`
[2] School of Engineering, Shahed University, Tehran, Iran
`ghaznavi@shahed.ac.ir` (*Corresponding Author)

**ABSTRACT**

*Preserving security and confidentiality in wireless sensor networks (WSN) are crucial. Wireless sensor networks in comparison with wired networks are more substantially vulnerable to attacks and intrusions. In WSN, a third person can eavesdrop to the information or link to the network. So, preventing these intrusions by detecting them has become one of the most demanding challenges. This paper, proposes an improved watchdog technique as an effective technique for detecting malicious nodes based on a power aware hierarchical model. This technique overcomes the common problems in the original Watchdog mechanism. The main purpose to present this model is reducing the power consumption as a key factor for increasing the network's lifetime. For this reason, we simulated our model with Tiny-OS simulator and then, compared our results with non hierarchical model to ensure the improvement. The results indicate that, our proposed model is better in performance than the original models and it has increased the lifetime of the wireless sensor nodes by around 2611.492 seconds for a network with 100 sensors.*

## 1. INTRODUCTION

The recent progresses in electronics and wireless telecommunication allow us to design and develop sensors with low consumptive power, small size, and reasonable price for various applications. These small sensors are able to receive different environmental information (based on the sensor type), process and transmit them.

Intruding into a network refers to any activity which endangers the integrity, confidentiality and accessibility of a source and an Intrusion Detection System (IDS) is a system which detects Intrusion activities [1]. The main idea of developing IDS came from examining the behavior patterns of ordinary users and identifying the abnormal behavior patterns of the users. Intrusion detection system which operates statistically demonstrates the network traffic like radar and detects any signal which may indicate an abnormal event or attack to the network [2].

Establishing security in wireless sensor networks due to its changing nature and non-concentrated typology has increased the vulnerability in these networks. Moreover, as a result of energy limitations in wireless sensors, the consumptive power has always been a challenging issue to be considered in designing these wireless sensor networks. In spite of the high volume of researches and studies which tried to propose an efficient intrusion detection system, none of them managed to develop a system which is able to identify and drive away all attacks to wireless sensor networks. Since, these researches have focused on specific kinds of attacks.

In this paper we, we propose an approach for implementing a new intrusion detection system to increase the network's lifetime and security level. The proposed algorithm solved the following known problems: impartial removal, selecting the incorrect malicious node, limited power transfer, and node conspiracy.

This paper organized in the following manner: Section 2 provides related works of intrusion detection in WSNs. Section 3, provides existing vulnerabilities of WSNs. The proposed algorithm and also the hierarchical model are also discussed in section 4. After that, the Tiny-OS simulation will be explained in Section 5. Experimental results are given in section 6. Section 7 concludes the paper and at the end, the future works will be discussed in section 8.

## 2. RELATED WORKS

So far, different techniques have been proposed for intrusion detection in wireless sensor networks. In the following sections some of them are discussed.

In [3] and [4] a technique has been proposed for identity authentication in intrusion detection in wireless sensor networks in an interleaved manner which is called Interleaved Hop-By-Hop authentication (IHOP in brief). IHOP guarantees to indentify all the incorrect packets injected into the network. In [3], the wireless sensors networks are organized hierarchically in clusters. The cluster in upper level creates a route for connection to base station and each interface node has a node connected to its upper level and also a node connected to its lower level. In IHOP an upper cluster collects the information related to identity authentication from its members (subordinates) and sends it to the base station in form of a report. This reporting occurs only when at least 1+t sensor observes similar results (this paper does not show how the t parameter should be adjusted to sensor network). However, IHOP guarantees that the base station will identify incorrect packets (when more than t nodes did not agree to cooperate).

Another technique [5] proposed route filtering using statistical techniques which can identify and delete incorrect data. In this technique, there is a key extensive pool and each sensor is allocated a part of this pool. Whenever a move in the region begins, the sensors identify this move and one of the nodes as the base station checks all the network addresses and filters all the reports en route conveying the address incorrectness. However, as mentioned in [4], this technique is used for protecting the network against incorrect information injection and cannot drive away the attacks such as selective forwarding attacks.

Also, another approach [6] was proposed based on a routing called INSESN (Intrusion-Tolerant Routing in Wireless Sensor Networks) in which the sensors collect the information related to regional typology and send it to the base station. Afterwards, the base station creates the routing table according to the collected information and sends it to the related sensors. The base station is the main control node for creating the routing table which reduces the nodes computational load. Although INSENS has been developed by a protocol based on routing table, these are the base stations which collect all the information and create the routing table for each sensor. However, INSENS is not suitable for large sensor networks.

During the recent years, intrusion detection based on the statistical techniques has been widely under the spotlight. For example, [7] uses data analysis techniques (such as clustering and neural networks [18]) using the data available in the user's reports and has examined and predicted their behavioral algorithm and tried to optimize the efficiency of intrusion detection in the system by separating the abnormal algorithms from the normal ones using various presuppositions and the intelligent technology.

## 3. SECURITY THREATS

The vulnerabilities in network cause attack occurrence. If we succeed to minimize them in a network, we can contribute to enhancing the security in the network. In fact, the vulnerabilities are the only controllable parts in wireless sensor networks. As a result of these natural limitations in WSNs, Denial Of Service (DOS) attacks which can cause damage to sensor networks by consuming the energy and the available sources [8] are the main power consumption attacks. Table 1 summarizes DOS attacks and the vulnerable areas which cause attack occurrence in WSN.

*Table 1. DOS Threats and Attacks in WSN*

| Attacks | Description |
| --- | --- |
| Jamming attack | Jamming attack occurs as a result of intentional interference in the radio waves in order to prevent a node from using radio channel [8]. |
| Tampering attack | Most of the professional attackers are able to intrude into the nodes memory and get access to the information inside or the encoding keys and Also, they can replace the programs with the malicious ones [9]. |
| Collision attack | In these attacks, the attacker identifies wireless exchanges around the victim node and creates Collision and destroys the key packets [8]. |
| Interrogation and Exhaustion attack | An attacker can create DOS attacks by inserting the nodes to his message re-sending [8]. |
| Selective forwarding attacks | In wireless sensor networks, all nodes can participate in routing operation to find the best route for sending the message. Only one node may not deny sending the packet and advertizing a specific route for its neighbors and it may create black holes in routing [10 & 11] |
| wrong-routing attacks (misdirecting attack) | An attacker can misdirect the network nodes by sending the messages to wrong routes [17]. |
| Sinkhole attacks | The attacker advertises wrong routing information [17]. |
| Wormhole attacks | In a wormhole attack, an attacker captures the packets at one point in the network and tunnels them to another point (at a distant location), and then replays them there into the network from that point [20]. |
| Sybil attacks | Most of the used protocols assume a unique ID. However in Sybil attacks, the attacker node has several IDs [12]. |
| Flooding Attack | Flooding attack covers and uses the victim's limited sources such as memory, processing cycle and band width [8]. |
| Hello Message Flooding | Since many protocols creates neighboring tables though exchanging Hello messages, these attacks cause various disorders in the system such as increasing the traffic in radio channels and decreasing the nodes operational power [13]. |

## 4. POWER-AWARE INTRUSION DETECTION IN WSNS

Power in the examined wireless sensor network can be obtained through computational techniques. Moreover, power as a desirable feature can prolong the life of sensor nodes and network [14]. Generally, the largest portion of the sensors power in wireless sensor network is consumed in receiving and transmitting the information at the RF module. Therefore, in this research, the main focus is on energy consumption in the functional areas of the wireless sensor network (such as the nodes containing IDS). $P_t(d)$, is the minimum power consumption for sending 1 bit of information at Euclidean distance d and $P_r$ is the minimum power consumed for receiving 1 bit which can by calculated by [14]:

$$P_t(d) = a_1 + a_2.d^n \tag{1}$$

$$P_r = B \tag{2}$$

Where $a_1$ the parameter related to sender circuit, equals 50 nj/bit, $a_2$ parameter related to sender booster, equals 100 Pj/bit/m$^2$, d stands for the distance between the functional node *i* and the target functional node, and *n* refers to the parameter related to the local emission reduction which equals 2. *B* refers to the parameters related to receiver circuit which is 50 nj/bit. Power consumption in a single node in time unit is calculated according to the equation 3 [14].

$$e_{it} = r_{ri}.P_r + (r_{ri} + r_{gi}).P_s \tag{3}$$

Where $r_{ri}$ refers to input information bit rate to the functional node i. $r_{gi}$ refers to the produced information bit rate in the functional node I and $P_s$ is minimum power consumed for sending information which is equal with $P_t(d)$. With regard to the fact that raw information bit rate in wireless sensor network is considerably low compared to the other common wireless networks, the high quality of the information and fast transfer is not of high importance. The average speed of raw information production is about 512 bps which is evidently very low in some applications. The life period of each functional node *i* equals to primary energy ratio to energy consumption in time unit. This is shown in the equation (4) [14]:

$$L_i = \frac{e_i}{e_{it}} \tag{4}$$

Where $e_i$ refers to the primary energy of the functional node *i* which is calculated according to:

$$e_i = R.I^2.t \tag{5}$$

According to [24], R stands for resistance which considered 1 ohm and I, is 8 mA for an active node, and t is the node's boot time.

### 4.1. Proposed Modified Watchdog Technique

Watchdog technique is one of the malicious nodes identification techniques which operates based on broadcast property in wireless sensor networks. The node like node A which intends to sends a packet to node C, can eavesdrop the sent traffic of the node B and determine whether or not the node B will send the packet to C (Figure 1).

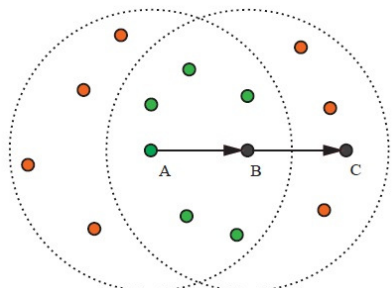

Figure 1. Conventional Watchdog mechanism

Here we try to propose a new technique based on Watchdog mechanism which is modified and improved by enhancing the security in wireless sensor network. We call this technique I-Watchdog (Improved Watchdog). Unlike the basic technique in which the node A is assumed to be the watchdog, in I-Watchdog technique, as shown in Figure 2, the cluster heads (nodes that are responsible for monitoring each cell) are assumed to be as the first layer watchdogs.

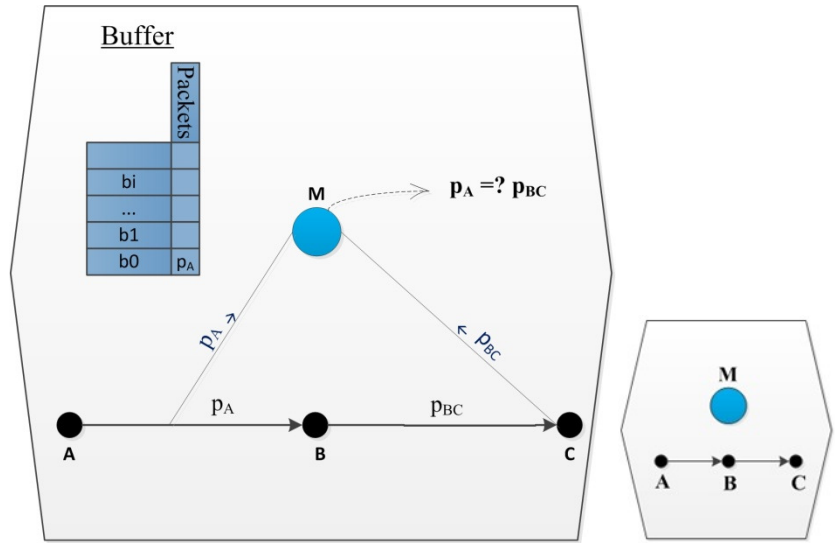

Figure 2. Proposed Improved-Watchdog (I-Watchdog)

In the proposed approach, if a node for example A wants to send a message to a node say C, the cluster head node (M) operates as watchdog. As shown in Figure 2, the cluster head node uses a buffer which accommodates all the sent items by the nodes within its sensory limit. Since the node B is an interface between A and C, it eavesdrops the first sending of the node B after receiving the message from A and compares it with the message in the buffer. If the messages are similar, the first message in the buffer will be deleted. Otherwise, it will turn out that the node B has not sent the message or replaced it with another one. We assume that the buffer used by Watchdog in this technique, as the Figure 2 show, has been divided into cells like $b_0$, $b_1$, $b_2$, …, $b_i$, …and $b_n$. According to malicious node detection algorithm which will be discussed below, and using the equation 6 value, we can determine whether or not the message sent from A to B will be correctly sent to C (the target node).

Here, $P_{BC}$ refers to the packet to be sent to C by B.

$$F_i = b_i - P_{BC} \qquad \forall \ 0 \leq i \leq n \tag{6}$$

### 4.1.1. The Proposed Algorithm for Malicious Node Detection

The steps of the algorithm for malicious node detection are as follow:

1- A sends a packet to C via B. Meanwhile, M (the cluster head node) eavesdrops the packet and saves a copy in its counterpart section in buffer b.

2- The node M eavesdrops to the communication between B and C for t second (this time depends on the nodes processing and sending speed as well as the sensors type) and refers to the step 5 in the case of not receiving any packet.
3- The $F_i$ value is calculated if M eavesdrops to the packet $P_{BC}$.
4- If $F_i=0$, the message in cell *bi* (where its counterpart message has been saved in buffer b) will be deleted and the algorithm moves to the step 6. If $F_i \neq 0$, the message remains in the buffer and moves to step 5.
5- The warning message, signaling the maliciousness of the node B, is sent to the upper layer by the cluster head node.
6- The end of algorithm.

As mentioned earlier, in the case of entering the step 5, the cluster head node sends a warning message to the upper layer. If the warnings reach a specific limit, the cluster head node introduces B and a malicious node. The flowchart of the conducted operation for intrusion detection has been illustrated in Figure 3 in order to facilitate understanding the proposed technique. In this flowchart, *i* is the buffer counter, $P_A$ is the packet sent by the node A, $P_{BC}$ is the sent packet from B to C, *t* is the maximum waiting time, *F* is the comparison function and *b(i)* is the content of cell *i* in buffer M.

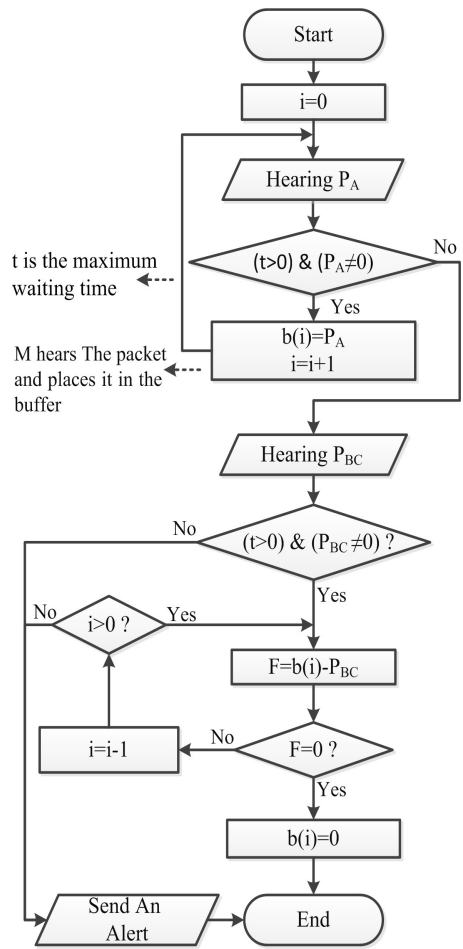

Figure 3. The Proposed flowchart for intrusion detection in WSNs

## 4.2. Error ratio

We can calculate the error ratio for selective forwarding attacks using the following equation, which has been mentioned in [15]:

$$Error\_ratio = \frac{bad\_couner}{bad\_couner + good\_counter} * 255 \qquad (7)$$

Where bad_counter is a number of packets that weren't forwarded and good_counter is a number of packets that were forwarded. The error ratio will help us to estimate mean error rate, so we can compare the proposed Technique with the original one.

## 4.3. Hierarchical Design Based on Intrusion Detection System

In this section, a model is proposed for saving the power consumed by the nodes while implementing an intrusion detection system in wireless sensor networks. This model follows a hierarchical architecture. The whole system is divided into smaller parts (called cells). Figure 4 shows the different layers categorization. Each cell indicates the sensory limit of a cluster head node. The cluster head nodes marked in blue are in charge of supervising the cells. According to this division, the red nodes are the regional nodes which as you can see should be selected in a way to be located on the cells boundaries to enable their sensory limit to cover a number of nodes of the cluster head.

It is important to be mentioned that, unlike the similar models such as [17], the proposed model does not necessarily require arranging the sensors in the system in order. The number of the sensors in the cells can be different. The system topology can be changing. The only fixed nodes in the network are the regional and cluster head nodes which should be selected at the outset of designing the network by the base station.

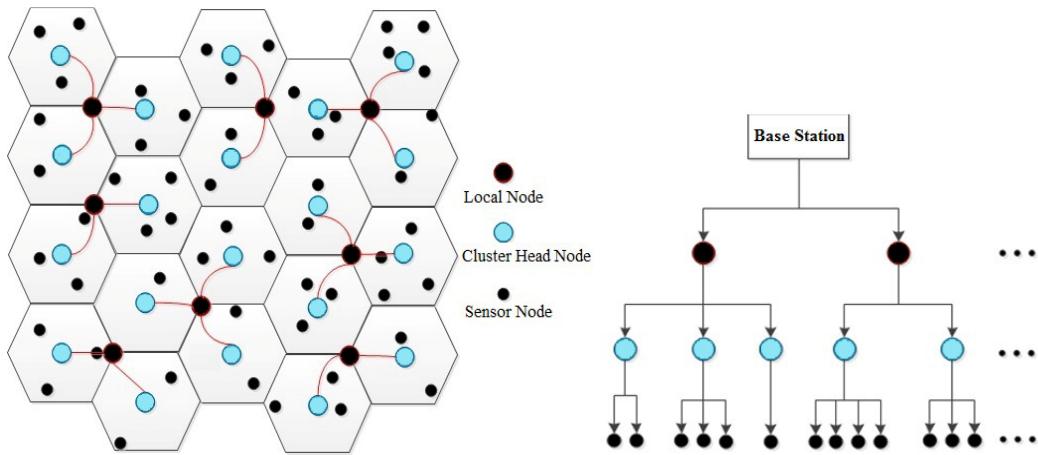

Figure 4. The modified hierarchical model and categorization of the entities

In the proposed model two intrusion detection mechanisms i.e. "signature-based detection" and "anomaly detection" can be incorporated. The normal attacks signatures can be compiled in the base station to be sent to the regional and cluster head nodes in order to detect these signatures. The date base is able to be updated to detect new attacks (detection based on the attacks signatures). In addition, using the techniques of detecting abnormality in the first level by the cluster head nodes, we can detect the attacks and hence create high security for the sensor network. The intrusion detection entities in the target architecture include:

*Cluster head node:* As mentioned earlier, this node is responsible for monitoring the related region. These nodes eavesdrop to the data sent by the nodes under their control, analyze the data and inform their upper nodes (regional nodes) of the suspicious cases. In fact, compared to the other sensors, these nodes enjoy more capacities including the intrusion detection program installed on them.

*Regional node:* These nodes are in charge of controlling and receiving the information from their neighboring cluster head nodes as well as sending warning message to the upper layer which is the base station. These nodes have all the abilities of the intrusion detection system. In addition, they allow integration into larger networks. Therefore, even in presence of a large number of sensors, the network can be divided into smaller section to be easily manageable.

*Base station:* Base station is the top level of the proposed model which is directly supported by human force. This station receives the information from the regional nodes, analyzes them and applies the necessary operations and policies to the system.

In the cases that message is sent to out of the cells limit, the monitoring is assigned to the upper layer. In this case, the regional nodes operate as Watchdog. If the regional nodes are malicious themselves, the upper layer i.e. the base station should detect the malicious nodes by monitoring the regional nodes and applying security measures. Thus, in this case the top level of watchdog is the base station. Figure 5 shows the different levels of watchdog in the network.

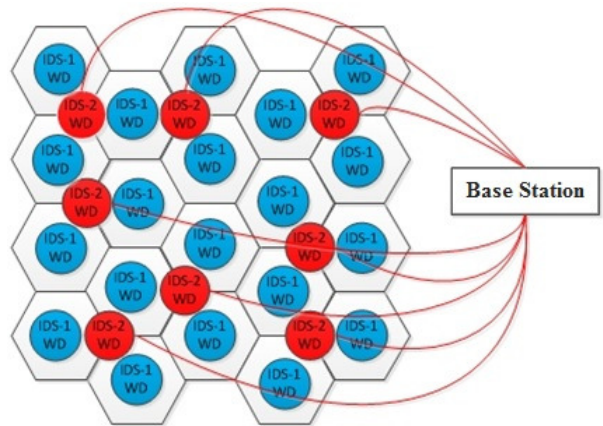

Figure 5. The Watchdog nodes in the proposed model

## 5. TINY-OS SIMULATION

In this paper Tiny-OS environment in Linux has been used for the simulations. TOSSIM is a Tiny-OS mote simulator which is useful for testing both the algorithms and implementations; however it does not simulate the physical phenomena that are sensed [19]. Tiny-OS is perhaps the first operating system specifically designed for wireless sensor networks [21]. Instead of compiling one of the Tiny-OS programs in each component of wireless sensor network, the users can use TOSSIM which can be run in PCs. Furthermore, it should be taken into the account that TOSSIM primarily aims at providing a simple and high-level simulation for Tiny-OS programs.

International Journal of Network Security & Its Applications (IJNSA), Vol.4, No.4, July 2012

Simulating sensor network have been conducted assuming 100 sensor nodes. Figure 6 indicates the lifetimes of the nodes which have different ranges in our simulated network. As shown in Figure 7, there are up to 40% of nodes which can live more than 60000 seconds, 21% of lifetimes are between 40000 to 60000 seconds and 30% of them are between 20000 to 40000 seconds.

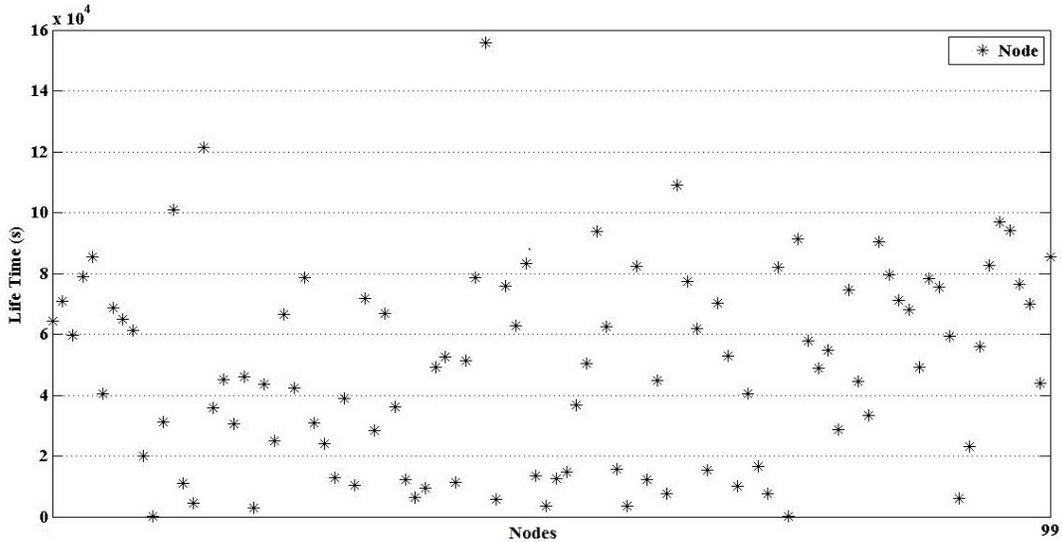

Figure 6. Nodes lifetime

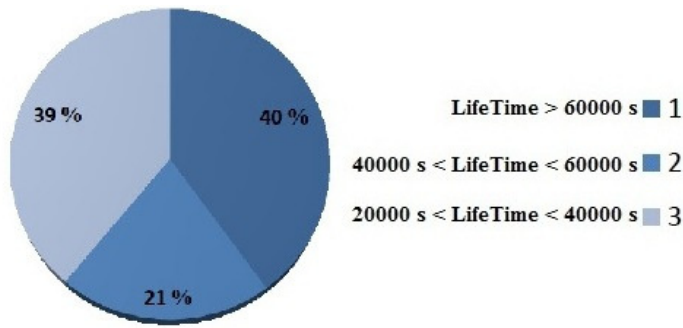

Figure 7. Share of nodes lifetime pie chart

Figure 8 shows the diagram of energy consumption variations by increasing the nodes distance (average radio signal). As you can see, the simulated network follows the theoretical pattern which implies that "the more the distance, the more the power consumption" and shows a similar behavior.

So, increasing the distance between nodes and the base station increases the consumed energy for sending and receiving the information and hence declines the node energy and remaining power. However, it is to be mentioned that other factors are also influential on nodes energy reduction such as primary energy of the nodes, the nodes sleeping cycle and the used intrusion detection system.

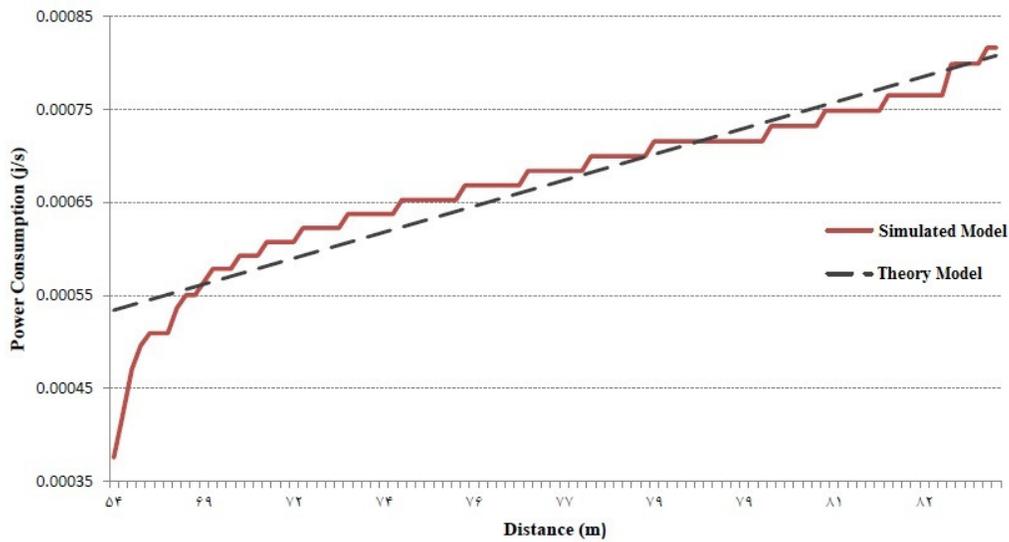

Figure 8. Diagram of nodes energy consumption variations by increasing the distance

With regard to the table 2, the proposed information can be interpreted as follows: the node 43 located in a 54 m distance from the base station has been booted with 58.53087 (Joule) primary energy and averagely (in active mode) consumes 0.00375 (Joule) energy per second and lives in the network for 156082.3 (seconds). Similarly, the node 83 located in a 79 m distance from the base station has been booted with 0.070435 (Joule) primary energy and averagely (in active mode) consumes 0.000716 (Joule) energy and will live in the system for 98.37274 (seconds).

Table 2. Minimum/ maximum power of the nodes

| Distance d (m) | The lifetime of the node i $L_i$ (s) | The energy consumed by the node $i$ in time unit $e_{it}$ (J/s) | Primary energy of the node $e_i$ (J) | Node's Number |
|---|---|---|---|---|
| 54 | 156082.3 | 0.000375 | 58.53087 | 43 |
| 79 | 98.37274 | 0.000716 | 0.070435 | 73 |

## 5.1. Cluster Head Selection

Cluster head nodes are selected according to their life period. Since we assume three-layer architecture, first we should select the high-level cluster head nodes then the low-level cluster head nodes. Since each cluster head node is in charge of monitoring the cell in two layers, we assume the number of the cluster head nodes in the second layer to be equal to the number of the cells. The number of the cells in a hierarchical network depends on several factors such as the number of the nodes, the area of the geographical environment and the sensors applications [22]. Since the number of the sensors is specified in the conducted simulation, this number is taken as the criteria for partitioning the system into cells. It is to be mentioned that, this layering can be different in other networks.

The simulation has been conducted for 100 sensor nodes. We assumed that if the maximum number of the possible nodes per cell is 5, then 20 cells can be put into the first layer. Similarly, if the maximum number of the possible nodes in the cells of the second layer is 4, then 5 cells can be put into the second layer. Therefore, the number of cluster head nodes is 25 (20 nodes in the first layer and 5 nodes in the second layer). Figure 9 shows this layering.

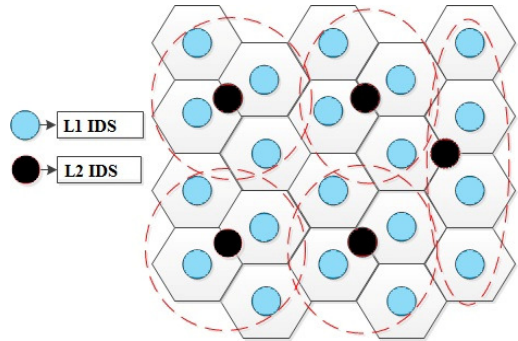

Figure 9. The layering of the wireless sensor network (assuming 100 sensor nodes)

As mentioned earlier, the cluster head nodes are responsible for collecting the information related to their cells and sending the information to the upper layer. In the hierarchical design, if the energy of some nodes (in a particular cell) got finished, the cluster head node can have access to the information through other sensors available in that cell. However, when the energy of the cluster head node ends, the base station's access to the cell's information will be lost, even if other sensors have energy in that cell.

**Threshold lifetime:** The cluster head node life should not be less than a specific time. We call it, threshold lifetime. This duration is different for each cluster head node due to the lower level nodes. It equals to the longest life period among the low-level sensors of a cluster head node.

## 5.2. Implementing IDSs

Implementing intrusion detection program on the designed wireless sensor networks requires measuring the energy consumed by the cluster head nodes for installing intrusion detection program on them. The cluster head node life duration should not be less than the low level nodes lifetime. Therefore, this power consumed for intrusion detection program should be considerably low so that the cluster head node life stays at the threshold life period level.

As mentioned before, in the proposed model, the intrusion detection programs can be only implemented on cluster head nodes or each cell representative. Therefore, the power consumed by the intrusion detection program should be also taken into the account in implementation process. Because some of the primary energy of the cluster head node $i$ is consumed for implementing the intrusion detection system while booting. According to energy survival law [23], the primary energy of the cluster head node $i$, by implementing the intrusion detection program once, can be calculated by below:

$$e_{ip} = e_i - e_p \tag{8}$$

Where $e_p$ is the power consumed to run the intrusion detection program on each node and $e_i$ is the primary energy of node $i$.

The wireless sensors network with consideration of the program's influence on cluster and regional nodes are shown in Figure 10. In this implementation, the location of the sensors is selected based on the distance from the base station.

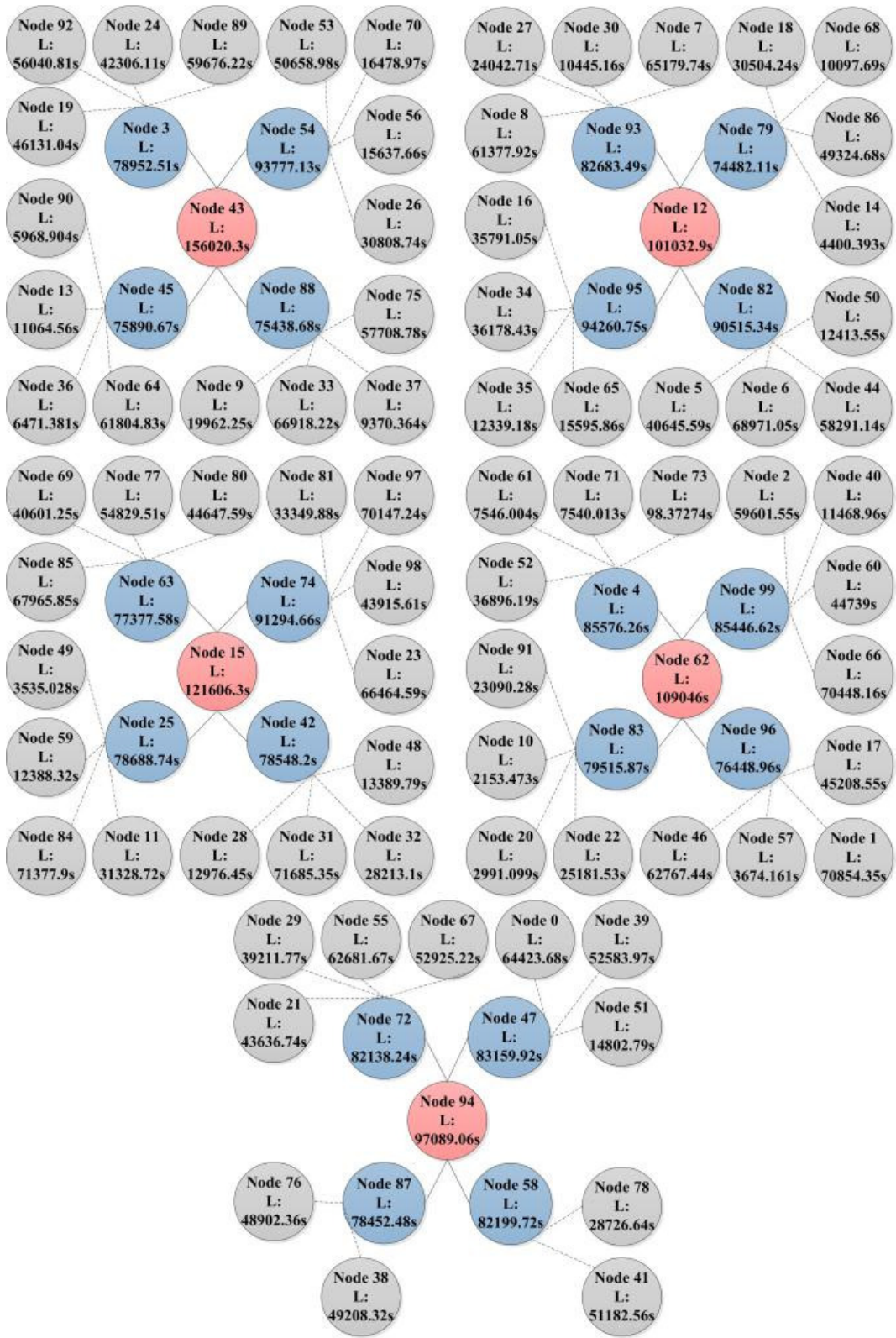

Figure 10. A view of our simulation setup with 100 nodes

## 6. EXPERIMENTAL RESULTS

In the first part of our experiments, we tried to set various sensing radius (-60 dB, -75 dB and -78 dB). The results of this test are shown on Figure 11. It is clear that when we limit monitoring only to nodes with stronger signal, mean error ratio is lower.

As can be seen in Figure 11, the improved Watchdog technique has less error than the original Watchdog technique and it seems to be more efficient. In addition, the highest mean error rates are within the -78 dB which is about 23 for I-Watchdog and 111 for Watchdog. According to our results, it appears that among these signals, the -60 dB is more appropriate.

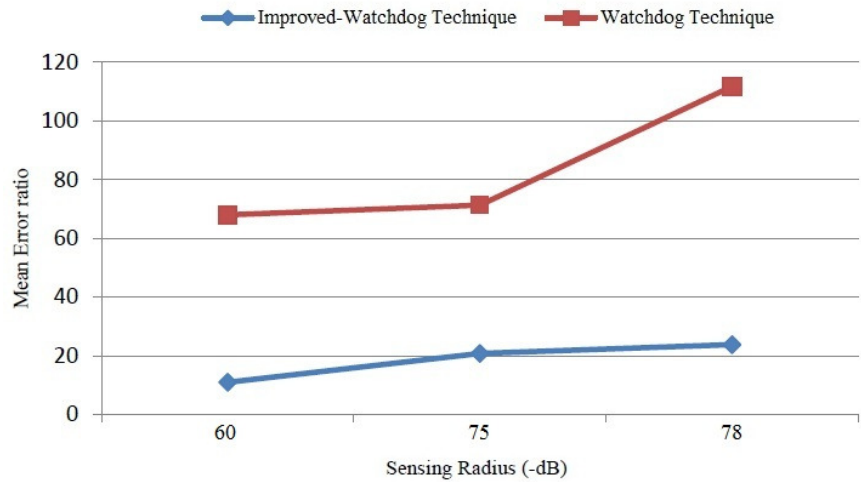

Figure 11. Dependence of error ratio on sensing radius

Table 3 shows the comparison between Watchdog and I-Watchdog techniques. As you can see, four problems of Watchdog techniques have been resolved in the proposed I-Watchdog technique.

Table 3. The comparison of I-Watchdog with Watchdog

| Problems | Watchdog | I-Watchdog |
|---|---|---|
| Creating ambiguous Collision | Yes | Yes |
| Creating Collision in the receiver | Yes | Yes |
| Selecting the incorrect malicious node | Yes | No |
| Limited power transfer | Yes | No |
| Node conspiracy | Yes | No |
| Impartial removal | Yes | No |

To comparing the network life period in the two model and normalize the nodes, we assume that the intrusion detection program has been already installed and implemented on all the nodes.

In the second part of our experiment, we proposed a hierarchical model; the intrusion detection program can be implemented only on the cluster head nodes (not on the other nodes). However, in the normal model, it can be implemented on all nodes and each node conducts the intrusion detection operation separately. Each node operates as a cluster head node. Figure 12 shows nodes life period difference in the proposed and the normal model.

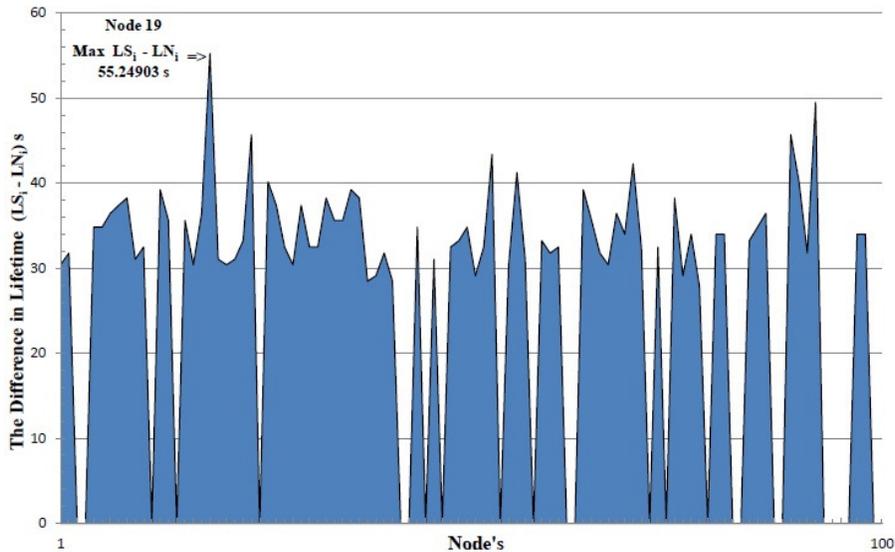

Figure 12. Nodes life period differences between the proposed and the conventional model

As seen in Figure 12, the biggest life period difference belongs to the node *19*. It means, this node (in the proposed model) lives 55.24903 seconds more than its counterpart node (in the normal model). Similarly, the other nodes are also indicated. According to Figure 12, it can be concluded that the proposed model has increased the sensor nodes lifetime approximately by *2611.492* seconds.

However, it is to be mentioned that in some of the nodes (e.g. node 3), life duration has not changed. Thus, in the proposed model, less energy is consumed for implementing the intrusion detection program. As a whole, in this model, 1744.488 mJ less energy was consumed compared to the normal model. This energy value is considerable for some wireless sensor network applications.

Finally, we simulated three Networks which were included 50, 100 & 200 nodes. After implementation of the proposed model, we have calculated the lifetimes of nodes, the overall network lifetimes with implementing the proposed model in compare to the normal model have been shown in Figure 13. As a result, there has been a rapid growth in the lifetime when the number of nodes increases.

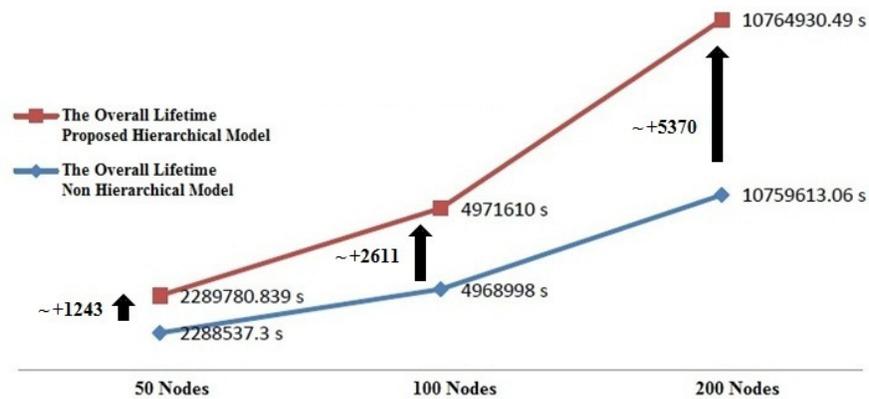

Figure 13. The overall network lifetimes by increasing the number of nodes

## 7. CONCLUSION

In this paper, we tried to prolong the sensor nodes lifetime by proposing a new hierarchical architecture and an improved version of Watchdog technique and implementing it. Also, we aimed at proposing an effective intrusion detection system in wireless sensor networks. The proposed model does not necessarily require arranging the sensors in the system in order and the number of the sensors in the cells can be changing. The system topology can also change. The only fixed nodes in the system are the regional and cluster head nodes which should be selected by the base station at the outset of network design. To confirm the optimization of this model compared to the normal sensor network model, the implementation has been conducted on a sensor network simulated in Tiny-OS. The results indicate that the proposed model is efficient in terms of energy consumption and sensor nodes life duration. In addition, the conclusions pertaining to the hierarchical model are as follow:

1- Network design based on layering technique enables the other sensors in a cell to cover the related region if the energy of one or more sensors ends. As a result, the sensor network operation will continue without encountering any problem.
2- Hierarchical design increases the network security. Because, if there is a malicious node in each layer, both the upper layer and the nodes in the same layer will detect it. Therefore, here the node maliciousness can be detected in two ways. Moreover, since the regional nodes are directly monitored by base station, they are not likely to be intruded.
3- Monitoring the layered networks is much more convenient. Because in the layered networks only the last layer nodes are monitored and the other nodes are managed by this layer. However, in the other networks, we are generally forced to be in touch with a wider range of nodes.

## 8. FUTURE WORK

Wireless sensor networks needs an intrusion detection system which operates regionally distributed. This system should be economical in terms of communications, energy and memory. So far, many studies have been carried out on establishing and preserving security and intrusion detection in wireless networks (such as wireless sensor networks). The salient research areas can be categorized as follows:

**Principles**: studies on intrusion, intruders and security threats.

**Information collection:** selecting the data sources and their features, how to collect the data, the staffs` entering and exit, system data format and template.

**Intrusion detection techniques:** finding best techniques for detecting intrusions, optimizing the intrusion detection efficiency.

**Reporting and reaction:** how inform the person or the network representative of the intrusion detection, how to react against attacks.

**Architecture of the intrusion detection system environment:** how distribute the intrusion detection agents in the environment and establish relation and interaction between them, also the issues related to IDS implementation on different systems and encoded networks.

**IDS security:** protecting IDSs and their traffic.

**Testing and evaluation:** how to test IDS and evaluate its performance.

**Operational aspects:** IDS maintenance, enhancement and transport.

**Social aspects:** issues related to preserving the user's privacy.

Most of the studies mentioned above have been reached an acceptable step. Most of them have been conducted on intrusion detection techniques.

## REFERENCES

[1] S. Şen, *et al.*, "Power-aware intrusion detection in mobile ad hoc networks," *Ad hoc networks,* pp. 224-239, 2010.

[2] Y. Wang, *"Statistical Techniques for Network Security: Modern Statistically-Based Intrusion Detection and Protection"* Idea Group Inc (IGI), 2008.

[3] S. Zhu, *et al.*, "Interleaved hop-by-hop authentication against false data injection attacks in sensor networks," *ACM Transactions on Sensor Networks (TOSN),* vol. 3, pp. 1-32, 2007.

[4] Y. Zhang and H. Hu, "*Security in wireless mesh networks*", Auerbach Publications, 2008.

[5] F. Ye, *et al.*, "Statistical en-route filtering of injected false data in sensor networks," *Selected Areas in Communications, IEEE Journal on,* vol. 23, pp. 839-850, 2005.

[6] J. Deng, *et al.*, "INSENS: Intrusion-tolerant routing for wireless sensor networks," *Computer Communications,* vol. 29, pp. 216-230, 2006.

[7] M. Khalilian, *et al.*, "Intrusion Detection System with Data Mining Approach: A Review," *Global Journal of Computer Science and Technology,* vol. 11, Issue 5, Version 1, pp. 1-7, 2011.

[8] A. D. Wood and J. A. Stankovic, "A taxonomy for denial-of-service attacks in wireless sensor networks," *Handbook of Sensor Networks: Compact Wireless and Wired Sensing Systems,* pp. 739-763, 2004.

[9] R. Anderson and M. Kuhn, "Tamper resistance: a cautionary note," *In Proceedings of the 2nd USENIX Workshop on Electronic Commerce*, Oakland, California, pp. 1-11, 1996.

[10] D. Ganesan, *et al.*, "Highly-resilient, energy-efficient multipath routing in wireless sensor networks," *ACM SIGMOBILE Mobile Computing and Communications Review,* vol. 5, pp. 11-25, 2001.

[11] S. Cheung and K. N. Levitt, "Protecting routing infrastructures from denial of service using cooperative intrusion detection," *Security Paradigms Workshop UK*, pp. 94-106, 1998.

[12]     J. Douceur, "The sybil attack," *Peer-to-peer Systems,* pp. 251-260, 2002.

[13]    C. Karlof and D. Wagner, "Secure routing in wireless sensor networks: Attacks and countermeasures," *Ad hoc networks,* vol. 1, pp. 293-315, 2003.

[14]    V. Gholampour, "Optimizing power consumption in wireless sensor network", M.A. thesis, Department of Computer and Electrical Engineering, Tehran University, 2003.

[15]    B. L. Honus, "Design, implementation and simulation of intrusion detection system for wireless sensor networks," Master Thesis, MASARYKOVA University, 2009.

[16]    P. K. Sree*, et al.*, "Power-aware hybrid intrusion detection system (PHIDS) using cellular automata in wireless ad hoc networks," *WSEAS Transactions on Computers,* vol. 7, pp. 1848-1874, 2008.

[17]    M. S. I. Mamun and A. F. M. S. Kabir, "Hierarchical Design Based Intrusion Detection System For Wireless Ad Hoc Sensor Network," *International Journal of Network Security & Its Applications (IJNSA),* vol. 2, no. 3,  pp. 102-117, 2010.

[18]    B. Shah and B. H. Trivedi, "Artificial Neural Network based Intrusion Detection System: A Survey," *International Journal of Computer Applications,* vol. 39, no. 6, pp. 13-18, 2012.

[19]    A. Dwivedi and O. Vyas, "An Exploratory Study of Experimental Tools for Wireless Sensor Networks," *Wireless Sensor Network,* vol. 3, pp. 215-240, 2011.

[20]    K. Nasr*, et al*, "Generating Representative Attack Test Cases for Evaluating and Testing Wireless Intrusion Detection Systems," *International Journal of Network Security & Its Applications (IJNSA),* vol. 4, no. 3, pp. 1-19, 2012.

 [21]    M. Bokare and M. A. Ralegaonkar, "Wireless Sensor Network: A Promising Approach for Distributed Sensing Tasks," *Excel Journal of Engineering Technology and Management Science*, vol. 1, pp. 1-9, 2012.

[22]    T.V. Padmavathy, *et al*, "Energy Constrained Reliable Routing Optimized Cluster Head for Multi hop under Water Acoustic Sensor Networks," *International Journal of Network Security & Its Applications (IJNSA),* vol. 4, no. 3, pp. 57-78, 2012.

[23]    I. F. Akyildiz*, et al.*, "Wireless multimedia sensor networks: A survey," *Wireless Communications, IEEE,* vol. 14, pp. 32-39, 2007.

[24]    Harvard University: UCB Mica2 Mote Power Benchmark Summary Numbers, URL:<http://www.eecs.harvard.edu/~shnayder/ptossim/mica2bench/summary.html> (view: 2012-03-02).

**Authors**

**A. Forootaninia** is a Master of Science student at the University of Tehran-Kish International campus in Iran. He received his Engineering Degree in Software engineering from the Islamic Azad University of Shiraz, Iran, in September 2009. He will receive his Master of Science degree from the University of Tehran-Kish international campus, Iran, in July 2012. His research interests include Information/Network Security (especially Wireless Sensor Networks). He may be reached at forootaninia@ut.ac.ir.

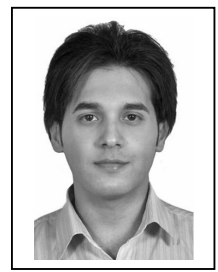

**M.B. Ghaznavi-Ghoushchi** received the B.Sc. degree from the Shiraz University, Shiraz, Iran, in 1993, the M.Sc. and Ph.D. degrees both from the Tarbiat Modares University, Tehran, Iran, in 1997, and 2003 respectively. During 2003-2004, he was a researcher in TMU Institute of Information Technology. He is currently an Assistant Professor with Shahed University, Tehran, Iran. His interests include VLSI Design, Low-Power and Energy-Efficient circuit and systems, Computer Aided Design Automation for Mixed-Signal and UML-based designs for SOC and Mixed-Signal.

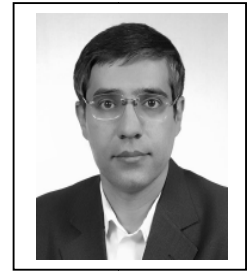